\documentclass[journal]{IEEEtran}
\usepackage{algorithm,amsbsy,amsmath,amssymb,epsfig,bbm,mathrsfs,multirow,amsthm}
\usepackage{array,multirow,graphicx}
\usepackage{graphicx}
\usepackage{graphics}
\usepackage{subcaption}
\usepackage{xcolor}
\usepackage{epstopdf}
\usepackage{color}
\usepackage{soul}
\usepackage{bbm}
\usepackage{cite}
\usepackage{algpseudocode}
\usepackage{mathtools}
\usepackage{pdfpages}
\usepackage{booktabs}

\newcommand{\beq}{\begin{equation}}
\newcommand{\eeq}{\end{equation}}
\newcommand{\bitm}{\begin{itemize}}
\newcommand{\ba}{\begin{array}}
\newcommand{\ea}{\end{array}}
\newcommand{\eitm}{\end{itemize}}
\newcommand{\beqn}{\begin{eqnarray}}
\newcommand{\eeqn}{\end{eqnarray}}
\newcommand{\beqno}{\begin{eqnarray*}}
\newcommand{\eeqno}{\end{eqnarray*}}
\newcommand{\bma}{\begin{displaymath}}
\newcommand{\ema}{\end{displaymath}}
\newcommand{\bnu}{\begin{enumerate}}
\newcommand{\enu}{\end{enumerate}}
\newcommand{\bce}{\begin{center}}
\newcommand{\ece}{\end{center}}
\newcommand{\btb}{\begin{tabular}}
\newcommand{\etb}{\end{tabular}}

\begin{document}
\title{Adversarial Attacks and Defenses in 6G Network-Assisted IoT Systems}

\author{Bui Duc Son, {Nguyen Tien Hoa,~\IEEEmembership{Member,~IEEE}}, Trinh Van Chien, \textit{Member}, \textit{IEEE},  Waqas Khalid, Mohamed~Amine~Ferrag, \textit{Senior~Member}, \textit{IEEE}, Wan~Choi, \textit{Fellow}, \textit{IEEE},  and Merouane~Debbah, \textit{Fellow}, \textit{IEEE} 


\thanks{Bui Duc Son and Nguyen Tien Hoa are with the School of Electrical and Electronic Engineering, Hanoi University of Science and Technology, Hanoi 100000, Vietnam. Emails:son.bd200524@sis.hust.edu.vn, hoa.nguyentien@hust.edu.vn.}

\thanks{Trinh Van Chien is with the School of Information and Communication Technology, Hanoi University of Science and Technology, Hanoi 100000, Vietnam. Email: chientv@soict.hust.edu.vn.}

\thanks{Waqas Khalid is with the Institute of Industrial Technology, Korea University, Sejong 30019, Korea. Email: waqas283@\{gmail.com, korea.ac.kr\}.}
\thanks{Mohamed Amine  Ferrag is with Technology Innovation Institute, 9639 Masdar City,
Abu Dhabi, United Arab Emirates. Email: mohamed.ferrag@tii.ae.}
\thanks{Wan Choi is with the Institute of New Media and Communications and the Department of Electrical and Computer Engineering, Seoul National
University (SNU),  Seoul 08826m Korea. Email: wanchoi@snu.ac.kr.}

\thanks{Merouane Debbah is with the Center for 6G
Technology, Khalifa University of Science and Technology,
Abu Dhabi 127788, United Arab Emirates Email: merouane.debbah@ku.ac.ae.}
}

\maketitle
\begin{abstract}
The Internet of Things (IoT) and massive IoT systems are key to sixth-generation (6G) networks due to dense connectivity, ultra-reliability, low latency, and high throughput. Artificial intelligence, including deep learning and machine learning, offers solutions for optimizing and deploying cutting-edge technologies for future radio communications. However, these techniques are vulnerable to adversarial attacks, leading to degraded performance and erroneous predictions, outcomes unacceptable for ubiquitous networks. This survey extensively addresses adversarial attacks and defense methods in 6G network-assisted IoT systems. The theoretical background and up-to-date research on adversarial attacks and defenses are discussed. Furthermore, we provide Monte Carlo simulations to validate the effectiveness of adversarial attacks compared to jamming attacks. Additionally, we examine the vulnerability of 6G IoT systems by demonstrating attack strategies applicable to key technologies, including reconfigurable intelligent surfaces, massive multiple-input multiple-output (MIMO)/cell-free massive MIMO, satellites, the metaverse, and semantic communications. Finally, we outline the challenges and future developments associated with adversarial attacks and defenses in 6G IoT systems.

\end{abstract}

\section{Introduction}
\label{sec:introduction}
Sixth-generation (6G) networks are considered a descendant of the current fifth-generation (5G) systems,  which is the next evolution of wireless technology \cite{8821934}. The expected benefits of 6G over previous generation networks include speed, data capacity, latency, reliability, and energy efficiency \cite{tariq2020speculative}. Specifically, the peak data rate in 6G communication is anticipated to be $1$ terabit per second (Tbps), approximately ten times faster than nowadays \cite{chen2020vision}. 6G systems are predicted to deliver substantially large data capacity, allowing more connected devices and data-intensive applications. This fact responds to the continuous increase of the Internet of Things (IoT) in the future years \cite{vaigandla2022communication,iote3}. Regarding latency, 6G systems are predicted to exceed the 5G networks by decreasing transmission delay to less than one millisecond (ms). Unprecedented low latency in 6G communication will allow new applications with real-time reactions and high accuracy, such as autonomous vehicles, remote surgery, and smart city applications.

Moreover, the upcoming evolution in cellular network technologies is expected to deliver more stable connectivity, with fewer dropped services and increased coverage in remote regions with harsh radio environments \cite{van2021reconfigurable}. The enhanced dependability should be vital for mission-critical applications, for instance, emergency services, industrial automation, and healthcare \cite{she2021tutorial}. Besides, 6G is predicted to be more energy-efficient than previous generations with the expectation of lowering power consumption and carbon footprint. This aim is to cope with the drastic demand for wireless access, and worries about climate change continue severely \cite{mukherjee2020energy}. On the other hand, security includes issues of paramount interest in 6G networks. There are several distinctions between canonical attacks and those of artificial intelligence-based networking. Canonical attacks in older generations such as 2G, 3G, and 4G mainly depend on exploiting vulnerabilities in communication protocols \cite{xu2017review}, network infrastructure\cite{wong2009network}, and user behavior \cite{peng2011security}. For example, many attacking techniques were used, such as denial-of-service (DoS) attacks \cite{boche2020denial}, A man-in-the-middle (MitM) attack \cite{bharati2020threats}, and malware attacks \cite{khouzani2012maximum} to disrupt network operations, steal sensitive data, and earn criminal access. \textcolor{black}{ 
Adversarial attacks includes new technologies where attackers intentionally manipulate input data to cause machine learning (ML) or deep learning (DL) models to make incorrect predictions or classifications. These manipulated inputs are known as adversarial examples. Specifically, we provide   Table~\ref{tab:my_label}, which summarizes the similarities and differences between canonical and adversarial attacks.} Even though both these attacks pose significant threats to wireless networks, adversarial attacks are more complex, constantly evolving, and presenting greater challenges to countering effectively. Because targets being attacked are vulnerable and attacks are difficult to detect, effective defenses to detect and minimize damages quickly are extremely important in practice. Such defenses should be dynamic and adaptive to respond to attack patterns by involving one or several advanced technologies, such as anomaly detection, adversary training, and behavioral analysis, to keep pace with the evolving threat landscape. 

\begin{table*}
\centering
\scriptsize
\caption{Compare between Traditional Attacks and Adversarial Attacks} \label{tab:my_label}
\begin{tabular}{|c|c|c|}
\hline
& Traditional Attacks  & Adversarial Attacks \\
\hline
Potentially compromise & \multicolumn{2}{|c|}{Confidentiality, integrity, and availability } \\
\hline
Exploit vulnerabilities & \multicolumn{2}{|c|}{Network, devices, protocols, or user behavior. } \\
\hline
Launched by & \multicolumn{2}{|c|}{Malicious actors} \\
\hline
Attacks require & \multicolumn{2}{|c|}{Network architecture, protocols,} \\
to be effective & \multicolumn{2}{|c|}{and security mechanisms} \\
\hline
Exploit & Known vulnerabilities & Evade detection or bypass defenses.\\
\hline
Scale & Generic network  & Specific network\\
\hline
Detected and mitigated & Traditional security mechanisms  & Advanced detection and mitigation techniques\\
\hline
Defense strategy & Static  & Dynamic and adaptive \\
\hline
Consequences & Damage or disruption to the network  & Compromising the trustworthiness of the network or the data it carries\\
\hline
\end{tabular}
\end{table*}

\begin{table*}[]
\scriptsize
\caption{Related Studies in Adversarial Attacks and Defenses for 6G-Assisted IoT Networks} \label{tab:studies}
\begin{tabular}{|l|l|l|l|l|lllll|}
\hline
\multicolumn{1}{|c|}{\multirow{2}{*}{Reference}} &
  \multicolumn{1}{c|}{\multirow{2}{*}{Year}} &
  \multicolumn{1}{c|}{\multirow{2}{*}{\begin{tabular}[c]{@{}c@{}}Centralized and \\ distributed attacks\end{tabular}}} &
  \multicolumn{1}{c|}{\multirow{2}{*}{\begin{tabular}[c]{@{}c@{}}Attack into \\ transmission phase\end{tabular}}} &
  \multicolumn{1}{c|}{\multirow{2}{*}{\begin{tabular}[c]{@{}c@{}}Attack into\\  resource allocation\end{tabular}}} &
  \multicolumn{5}{c|}{Attack into new technology use in 6G} \\ \cline{6-10} 
\multicolumn{1}{|c|}{} &
  \multicolumn{1}{c|}{} &
  \multicolumn{1}{c|}{} &
  \multicolumn{1}{c|}{} &
  \multicolumn{1}{c|}{} &
  \multicolumn{1}{c|}{RIS} &
  \multicolumn{1}{c|}{\begin{tabular}[c]{@{}c@{}}Massive MIMO  \\ and Cell-free \\ Massive MIMO\end{tabular}} &
  \multicolumn{1}{c|}{\begin{tabular}[c]{@{}c@{}}Space-ground \\ comm.\end{tabular}} &
  \multicolumn{1}{c|}{Metaverse} &
  \multicolumn{1}{c|}{\begin{tabular}[c]{@{}c@{}}Semantic \\ comm.\end{tabular}} \\ \hline
\cite{9060970} &
  2020 &
  no &
  no &
  yes &
  \multicolumn{1}{l|}{no} &
  \multicolumn{1}{l|}{no} &
  \multicolumn{1}{l|}{no} &
  \multicolumn{1}{l|}{no} &
  no \\ \hline
\cite{9482503} &
  2021 &
  no &
  no &
  no &
  \multicolumn{1}{l|}{no} &
  \multicolumn{1}{l|}{no} &
  \multicolumn{1}{l|}{no} &
  \multicolumn{1}{l|}{no} &
  no \\ \hline
\cite{9524814} &
  2021 &
  no &
  yes &
  yes &
  \multicolumn{1}{l|}{yes} &
  \multicolumn{1}{l|}{yes} &
  \multicolumn{1}{l|}{\begin{tabular}[c]{@{}l@{}}no (attack  \\ not detailly \\ discussed)\end{tabular}} &
  \multicolumn{1}{l|}{no} &
  no \\ \hline
\cite{9509294} &
  2021 &
  yes &
  yes &
  \begin{tabular}[c]{@{}l@{}}no (attack \\  not detailly\\  discussed)\end{tabular} &
  \multicolumn{1}{l|}{\begin{tabular}[c]{@{}l@{}}no (attack  \\ not detailly \\ discussed)\end{tabular}} &
  \multicolumn{1}{l|}{\begin{tabular}[c]{@{}l@{}}no (attack  \\ not detailly \\ discussed)\end{tabular}} &
  \multicolumn{1}{l|}{yes} &
  \multicolumn{1}{l|}{no} &
  no \\ \hline
\cite{10263803} &
  2022 &
  \begin{tabular}[c]{@{}l@{}}no (Centralized \\ not detailly \\ discussed)\end{tabular} &
  \begin{tabular}[c]{@{}l@{}}no (channel \\ estimation not \\ detailly discussed)\end{tabular} &
  yes &
  \multicolumn{1}{l|}{yes} &
  \multicolumn{1}{l|}{yes} &
  \multicolumn{1}{l|}{no} &
  \multicolumn{1}{l|}{no} &
  no \\ \hline
\cite{9711564} &
  2022 &
  \begin{tabular}[c]{@{}l@{}}no (attack  \\ not detailly\\  discussed)\end{tabular} &
  \begin{tabular}[c]{@{}l@{}}no (attack  \\ not detailly \\ discussed)\end{tabular} &
  \begin{tabular}[c]{@{}l@{}}no (attack\\  not detailly \\ discussed)\end{tabular} &
  \multicolumn{1}{l|}{no} &
  \multicolumn{1}{l|}{\begin{tabular}[c]{@{}l@{}}no (attack \\  not detailly\\  discussed)\end{tabular}} &
  \multicolumn{1}{l|}{\begin{tabular}[c]{@{}l@{}}no (attack \\ not detailly \\ discussed)\end{tabular}} &
  \multicolumn{1}{l|}{no} &
  no \\ \hline
\cite{9919765} &
  2022 &
  \begin{tabular}[c]{@{}l@{}}no (Centralized \\  not detailly \\ discussed)\end{tabular} &
  yes &
  yes &
  \multicolumn{1}{l|}{no} &
  \multicolumn{1}{l|}{no} &
  \multicolumn{1}{l|}{yes} &
  \multicolumn{1}{l|}{no} &
  no \\ \hline
\cite{10257196} &
  2023 &
  no &
  no &
  no &
  \multicolumn{1}{l|}{no} &
  \multicolumn{1}{l|}{no} &
  \multicolumn{1}{l|}{no} &
  \multicolumn{1}{l|}{no} &
  no \\ \hline
\cite{10181170} &
  2023 &
  yes &
  no &
  no &
  \multicolumn{1}{l|}{no} &
  \multicolumn{1}{l|}{no} &
  \multicolumn{1}{l|}{no} &
  \multicolumn{1}{l|}{no} &
  no \\ \hline
\cite{10255264} &
  2023 &
  yes &
  yes &
  yes &
  \multicolumn{1}{l|}{\begin{tabular}[c]{@{}l@{}}no (attack \\  not detailly \\ discussed)\end{tabular}} &
  \multicolumn{1}{l|}{\begin{tabular}[c]{@{}l@{}}no (attack \\  not detailly \\ discussed)\end{tabular}} &
  \multicolumn{1}{l|}{\begin{tabular}[c]{@{}l@{}}no (attack \\ not detailly \\ discussed)\end{tabular}} &
  \multicolumn{1}{l|}{no} &
  \begin{tabular}[c]{@{}l@{}}no (attack \\  not detailly \\ discussed)\end{tabular} \\ \hline
Our survey &
  2024 &
  yes &
  yes &
  yes &
  \multicolumn{1}{l|}{yes} &
  \multicolumn{1}{l|}{yes} &
  \multicolumn{1}{l|}{yes} &
  \multicolumn{1}{l|}{yes} &
  yes \\ \hline
\end{tabular}
\end{table*}

\begin{figure}[t]
\includegraphics[trim=0.0cm 0.0cm 0.4cm 0.4cm, clip=false, width=3.4in]{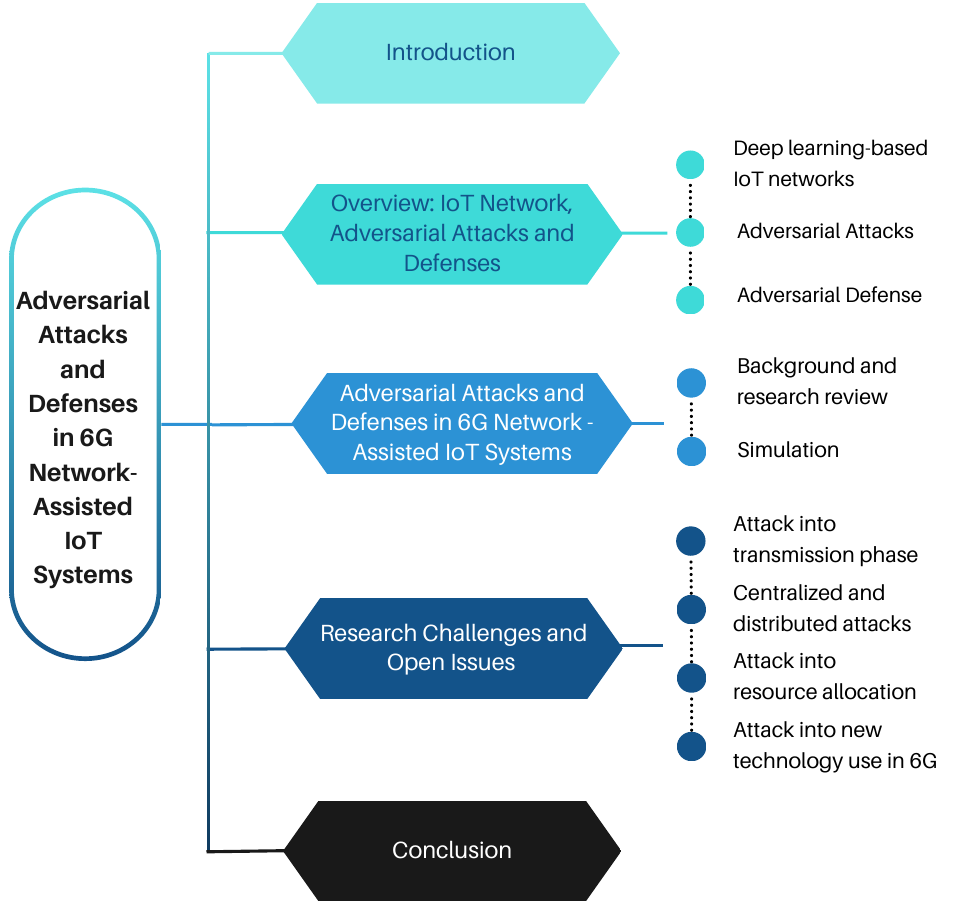}
 \caption{Organization of the paper}
  \label{Fig:Organization}
\end{figure}

The recent development of artificial intelligence with the support of hardware innovation, more and more new attack methods have appeared nowadays. The efficiency and effectiveness of DL and ML demonstrate their roles in designing high-performance systems. However, the behaviors of learning-based models may be influenced by the poison of malicious data so-called adversarial attacks\cite{ren2020adversarial}. These new attacking technologies are capable of obstructing communication reliability \cite{ftaimi2021towards}, violating user privacy \cite{zhang2020adversarial}, and stealing data \cite{liu2020privacy}. To control resources and enhance network performance, 6G IoT wireless networks must be robust to vulnerabilities produced by adversarial attacks, resulting in the so-called adversarial defenses\cite{wong2018scaling}.  Defensive algorithms may be generally divided into reactive and proactive categories. In particular, proactive adversarial defensive mechanisms intend to stop assaults before they have ever started. In contrast, adversarial reactive defense mechanisms aim to identify and reduce hostile attacks after they have occurred.  Due to the fast growth of IoT devices, advanced modern networks, and cutting-edge artificial intelligence applied for communication engineering, it calls for an extensive overview of adversarial attacks and defenses for the upcoming IoT systems towards 2030. In particular, our main contributions are summarized as follows:
\begin{itemize}
\item We propose an overview approach for using adversarial attacks and defenses in the context of 6G network-assisted IoT systems. Adversarial attacks and defenses are introduced with definitions, classifications, and applications.  Additionally, specific architectures and performance potentially applied for IoT systems are discussed.
\item We consider the opportunities and approaches for adversarial attacks and defenses in various parts of 6G based on the literature review and the theory's background. The opportunities for adversarial attacks and their possible defenses in data transmission, the density of equipment, resource allocation, and the new technologies in 6G IoT networks are demonstrated to be robust and effective with the support of ML and DL.
\item We compare adversarial attacks with canonical jamming attacks to the auto-encoder systems using the block error ratio under the severity of noise plus interference. The practical scenarios aim to comprehend better how auto-encoders perform in the face of these attacks.
\item Finally, we present Table~\ref{tab:studies}, which provides a detailed comparison between our work
and state-of-the-art studies.
\end{itemize}
The organization of this paper is illustrated in Fig.~\ref{Fig:Organization}. The overview of adversarial attacks and defenses, in general, is given in Section~\ref{sec:system_model}. Adversarial attacks and defenses, in particular for 6G network-assisted IoT systems, including background, research review, and simulation are presented in Section~\ref{Sec:AdverDef}. Finally, Section~\ref{Sec:ResearchChallegnes} presents the attacks and defense approaches in 6G network-assisted IoT systems and the main conclusions are drawn in Section~\ref{sec:conclusion}.

\section{Overview: IoT Network, Adversarial Attacks and Defenses}
\label{sec:system_model}
This section provides a comprehensive overview of ML-based adversarial attacks and defenses with definitions, classifications, architectures, performances, and the effectiveness of these techniques.
\subsection{Deep learning-based IoT networks}

In IoT networks, electronic devices with software and hardware are interconnected in a coverage area. Over the last decade, IoT devices have increased significantly, resulting in enormous data transmission. Due to the limited resources, such as the power budget and system bandwidth, resource allocation plays a significant role in optimizing and improving network performance. Fortunately, the introduction and development of DL and ML are suitable for planning and managing these limited resources. Specifically, neural networks are applied in many fields in IoT networks, including smart homes and medical, industrial, and environmental sectors. In the range of this paper, we focus on the physical layer that includes potentially advanced technologies for 6G communications. We concentrate on how ML and DL operate for IoT components like sensors and other embedded devices that respond to transmit, receive, and capture data with finite radio resources.
\begin{figure}[t]
\includegraphics[trim=0.0cm 0.0cm 0.4cm 0.4cm, clip=false, width=3.4in]{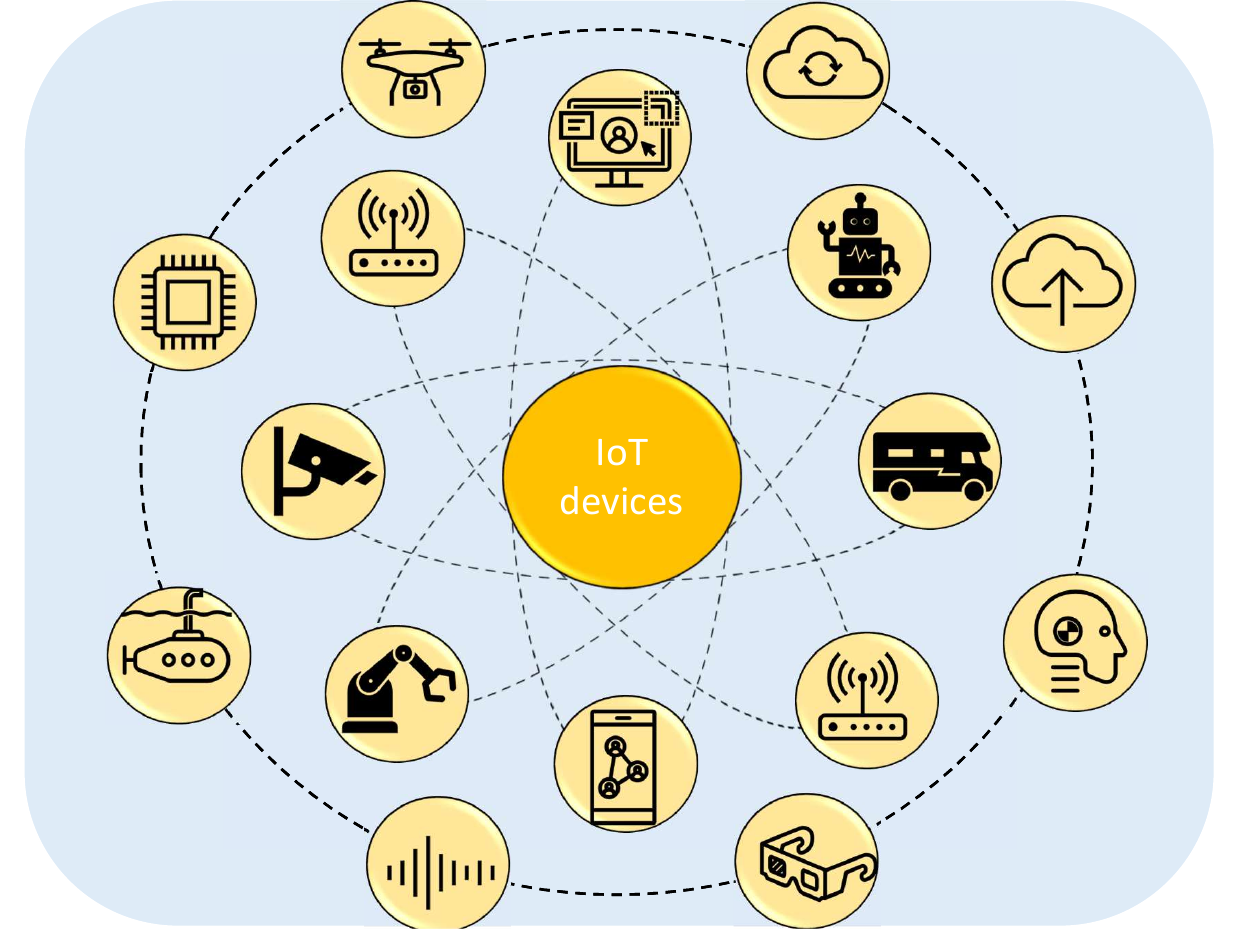}
 \caption{IoT devices}
  \label{Fig:IOT}
\end{figure}
Moreover, IoT devices responsible for collecting and transmitting data from the physical world include sensors, cameras, and actuators in various applications, as shown in Fig.~\ref{Fig:IOT}. Due to the massive connectivity, the volume of data required for processing is enormous. Hence, it calls for efficient techniques to manage all the activities by inheriting hardware development. Although IoT devices would have different sizes, shapes, and types depending on the functionalities and applications, modern neutral networks should be adapted to handle particular scenarios. The IoT components can also be seen as standalone devices using wireless communication protocols. Therefore, DL and ML can act as anomaly data detection or classification data. Moreover, embedded systems can be considered the brains of IoT networks that play an important role in controlling spectral and energy efficiency. Hence, we exploit artificial intelligence to inherit its flexibility and adaptability to achieve the best performance. Because of the limitations in microcontrollers, microprocessors, memory, storage, operating systems, and firmware, storage is limited. DL and ML have, therefore, significant roles in allocating resources effectively with low-cost operators. We stress that the trade-off of using modern learning-based methods is computational complexity and prediction performance. So, advanced learning techniques are promising for enhancing IoT networks. Additionally, with the limitations of both hardware and radio resources, intelligent strategies to allocate these resources efficiently are necessary for that kind of network, which is related to maximizing the profits and performance of that network. The IoT reliability determines its equipment working with high performance meaning that solving the optimization problems is expected to obtain effective solutions. Because of the variations of a physical environment, IoT components may switch the hardware levels in different time slots to catch real-time implementation. Thus, DL models are appropriate for this mission. In more detail, the authors in \cite{10064048} proposed DL architectures that minimize the transmitted power. During the testing phase, the proposed neural networks balanced between the energy consumption and the rate limitations for the enormous connectivity of IoT devices. Moreover, X. Liu et al. \cite{liu2020resource} suggested the ML approaches for edge computing-assisted IoT networks. Specifically, by employing the deep-Q network, the networks were able to optimize radio resources and discover the optimal offloading policies. The results showed that the proposed algorithms were more effective than the other benchmarks over the same network settings.

\subsection{Adversarial Attacks}
\subsubsection{Definition} 
In this part, we will provide the basic knowledge of adversarial attacks. An adversary is an unknown object or person who wants to perturb clean data and cause trouble using ML or DL models. Adversarial perturbation is a way that traps learning models to make wrong decisions. In addition, an adversarial example is a data set containing adversarial perturbations to fool and pose a significant threat to learning-based systems. This attack can target vulnerabilities in various modern learning systems, including deep neural networks (DNNs), convolutional neural networks(CNNs), long short-term memory, and transformers.  Each threat has the main objective of deceiving the model by manipulating the input data to make incorrect decisions, leading to significant damage such as decreasing effectiveness. We now classify adversarial attacks.

\subsubsection{Classifications}
Adversarial attacks can be classified based on the adversary's knowledge, methods, goals, and attacking phase.
First, we consider the categorization based on attacking mechanisms, including exploratory attacks (inference attacks), evasion attacks, causative attacks, and trojan attacks. In particular, inference attacks \cite{lin2020exploratory} search the habits and activities of cutting-edge learning models or algorithms for a given data set. Note that inference attacks are often used in the first phase to build a clone model for testing the performance of attack approaches. Meanwhile, Evasion attacks \cite{8439941} modify input data in a way that can fool DL models without being detected by humans. It can lead to incorrect decisions and reduce the reliability of neural networks by perturbing the input data. Causative attacks \cite{shi2017evasion} focus on damaging neural networks from the training phase, to manipulate the training data. Trojan attacks \cite{bhunia2014hardware} indicate the cooperation between the evasion and causative attacks to inherit the advantages of both attacks. The attacker injects triggers (backdoors) into training data and then activates them to misclassify the input in the testing phases.

\begin{figure*}[t]
\centering
\includegraphics[width=0.8\linewidth]{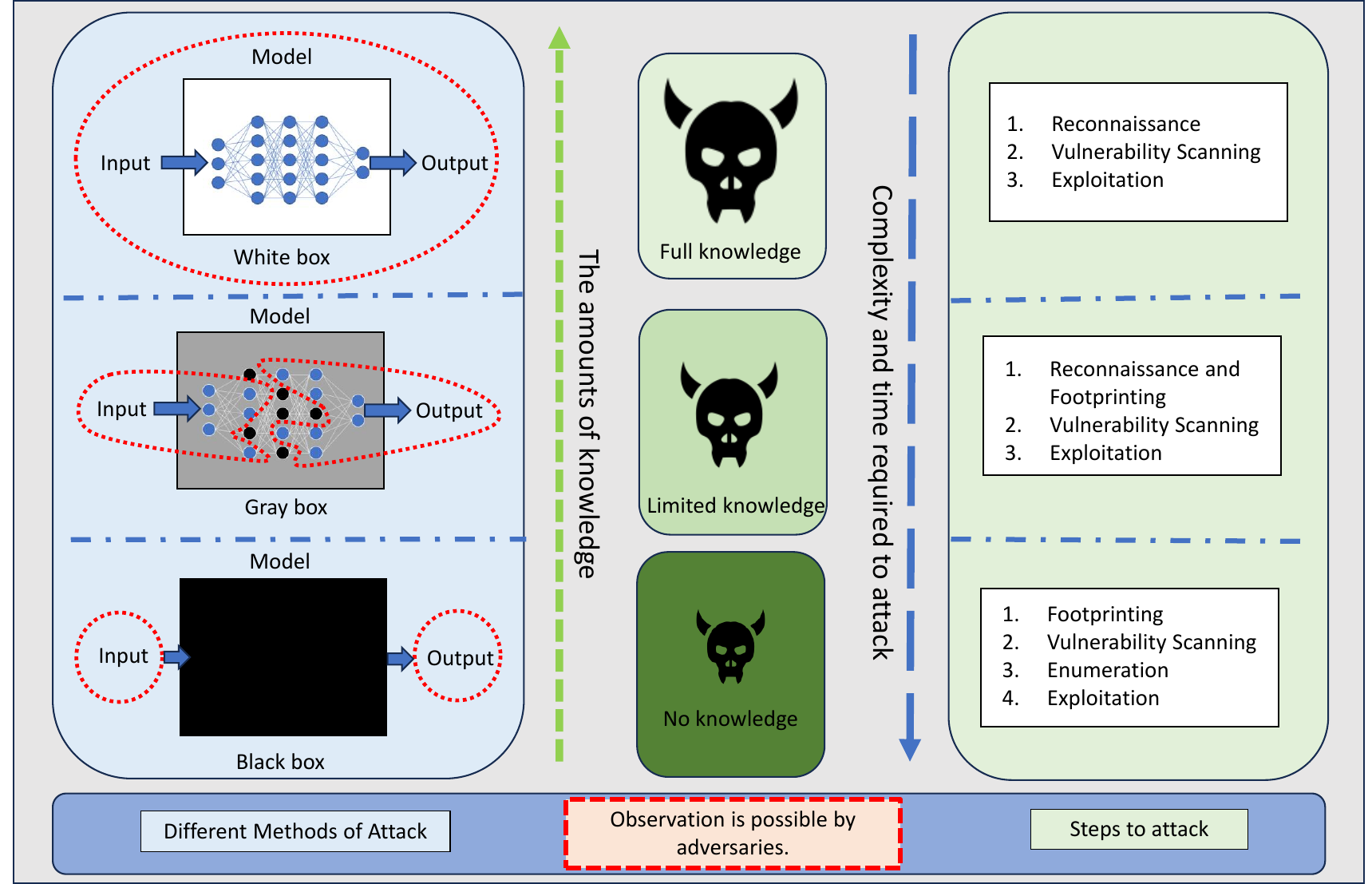}
 \caption{Adversarial attacks based on the adversary's knowledge of the victim models.}
\label{fig:3}
\end{figure*}

In another classification, we can sort out the attacks based on the adversary's knowledge of the victim models, which consist of the black box, white box, and gray box attacks. In the black box attack \cite{papernot2017practical},  the attacker attempts to reconstruct a mirrored model version without having access to its construction details. Specifically, the attacker can only observe the input and output of a model and tries to imitate the same task as the target model using an inference. Oppositely, the white box attacks \cite{nasr2019comprehensive} capture the targeted model's knowledge, including the input, output, architecture, algorithm, and optimization techniques. In most cases, the attackers can access the targeted model. Combination between the black and white box attacks are the gray box attacks\cite{vivek2018gray}. Since the adversary understands the model partially, attacking has some explicit limitations. This categorization is illustrated in Fig.~\ref{fig:3}.

\label{summary1}
Besides the main categorizations mentioned above, there are a few distinctive classifications, such as assortment based on adversary’s goals (targeted \cite{10147340} and non-targeted \cite{9706929}). After that, we introduce some famous attack architectures of the attack methods hereafter.

\subsubsection{Methodology of the attacks}
In an attacker's vision, selecting a proper attacking method is paramount to degrade the system's reliability under the given transmission conditions. We will describe the attacks' methodology based on the knowledge required to violate a system.  As previously mentioned, in black box attacks, adversaries lack knowledge of the architecture of the system.  They only rely on the input and output information to interfere and ruin a network, as shown in Figure \ref{fig:3}. In real-world scenarios, attackers often lack the details of the victim models. To determine the effectiveness of the architecture, we will be using a black box attack. This is because it is a feasible approach to take. The black box attack consists of four steps, and there are four main methods that we will be examining.

\begin{algorithm}[t]
\caption{Fast Gradient Sign Method (FGSM)}
\label{alg:FGSM}
\begin{algorithmic}[1]
\Require Input $x$, target model $f$, perturbation budget $\epsilon$
\Ensure Perturbed input $x_{\text{perturbed}}$
\State Compute the gradient of the loss function $J$ with respect to $x$:
\State $\quad \quad \quad \quad \quad$ $\text{gradient} = \nabla_x J(f(x), y_{\text{true}})$, where $y_{\text{true}}$ is the true label of $x$
\State Compute the sign of the gradient:
\State $\quad \quad \quad \quad \quad$ $\text{sign} = \text{sign}(\text{gradient})$
\State Create the perturbed input by adding the scaled sign of the gradient to the original input:
\State $\quad \quad \quad \quad \quad$ $x_{\text{perturbed}} = x + \epsilon \cdot \text{sign}$
\State Use the perturbed input $x_{\text{perturbed}}$ to make a prediction with the target model $f$
\State \Return $x_{\text{perturbed}}$
\end{algorithmic}
\end{algorithm}

\begin{algorithm}[t]
\caption{Projected Gradient Descent (PGD)}
\label{alg:pgd}
\begin{algorithmic}[1]
\Require Input $x$, target model $f$, step size $\alpha$, maximum number of iterations $K$, perturbation budget $\epsilon$
\State Initialize $x_0 = x$
\For{$k=1$ to $K$}
\State Compute the gradient of the loss function $J$ with respect to $x_k$:
\[
\text{gradient} = \nabla_x J(f(x_k), y_{\text{true}})
\]
\State Compute the sign of the gradient and add noise to it:
\[
\widetilde{\text{gradient}} = \text{sign}(\text{gradient}) + \mathcal{U}[-\epsilon,\epsilon]
\]
where $\mathcal{U}[-\epsilon,\epsilon]$ is uniform noise in the range $[-\epsilon,\epsilon]$.
\State Project the perturbed input $x_k + \alpha \widetilde{\text{gradient}}$ back onto the $\ell_\infty$-norm ball of radius $\epsilon$ around the original input $x$:
\[
x_{k+1} = \text{clip}_{x,\epsilon}(x_k + \alpha \widetilde{\text{gradient}})
\]
where $\text{clip}_{x,\epsilon}(x')$ is a function that clips each element of $x'$ to be within $\epsilon$ of the corresponding element of $x$.
\EndFor
\State Use the perturbed input $x_K$ to make a prediction with the target model $f$.
\State The final perturbed input $x_K$ will likely be misclassified by the target model $f$, even though it is very similar to the original input $x$.
\end{algorithmic}
\end{algorithm}
First, we clarify the methodology of the black box based on the transfer attack \cite{cheng2019improving}, which is defined by a synthesis of the corresponding white box attacks.  In more detail, the adversary can observe the training data and then try to simulate their features and validate various test cases of the white box attacks on the trained model to find the vulnerability and then damage the targeted victim. One of the popular algorithms used in adversarial attacks is the fast gradient sign method (FGSM) \cite{9682097}. FGSM uses the neural network's gradients to create adversarial examples as presented in Algorithm~\ref{alg:FGSM}. Another effective method, the projected gradient descent (PGD) algorithm, implemented in Algorithm \ref{alg:pgd}, is also utilized in this attack. Note that both FGSM and PGD add small perturbations. However, PGD iteratively perturbs the training data and ensures that the adversarial examples are well-suitable by projecting them back onto a good input space. In some complicated scenarios, more advanced algorithms such as the Carlini and Wagner (CW) attack \cite{chen2020stateful} and the Jacobian-based Saliency map attack (JSMA) \cite{liu2021practical} can be applied for higher attacking performance than the two previous mentioned methods.

Regarding the black box based on the reward attack \cite{cai2022black}, the adversary aims to estimate the reward of the targeted model to create optimized adversarial examples. The following are the general steps of a black box based on the reward attack: $i)$ given input $x$ and target model $f$, generate a generative model $g_\theta$ that can be trained to create samples similar to $x$; $ii)$ train the generative model $g_\theta$ with a set of clean data points, using the target model $f$ as a reward signal; $iii)$ after completing the training phase of the generative model $g_\theta$, use it to create various samples; $iv)$ select the most suitable sample that maximizes the reward of the target model $f$; and $v)$ the selected sample as an adversarial example to attack the target model $f$. The main advantage of the reward attack is that it is effective against black-box models without requiring knowledge of the target model's internal design. Moving on, another type of attack focuses on the decision-making process, the black box attack based on the decisive attack \cite{brendel2017decision}. Notably, the decisive attack does not require any reward vector in the output of the synthetic model. It needs a hard label output so the classifier cannot detect testing data correctly.
The last classification is the black box based on the non-traditional attack \cite{salamh2019drone}. Unlike the approaches mentioned before, this classification is more general. We stress that some black-box attacks do not fit into the previously mentioned attacks.
However, the landscape of those attacks is extensive and not thoroughly covered here. To delve deeper into the topic and obtain the best performance and analysis, it is advisable to refer to dedicated research papers such as \cite{salamh2019drone}, \cite{wang2022black}, \cite{alatwi2021adversarial}, and others.
In addition to the black box attack, we also consider the white box \cite{10145466} and gray box attacks \cite{9970367}. The white box attack involves knowing the model's architecture, which makes it more flexible than the black box attack. Essentially, the white box attack encompasses the architecture of the black box and has significant knowledge or "power". On the other hand, the gray box attack shares the same architectural framework but lacks the strong influence of the white box. In conclusion, both white box and gray box attacks have superior architecture and are more effective in targeting ML and DL models. However, these approaches may not be practical in cases where we lack any knowledge about the target model, making them less realistic.

\subsubsection{Performances and effectiveness} 
Performance and energy efficiency are key in certain scenarios. In this research, we evaluate the effectiveness of black-box attacks, commonly used to test these scenarios.

The authors in \cite{papernot2016transferability} demonstrated the efficacy of black-box attacks on commercial ML classifiers from Amazon and Google, achieving misclassification rates of 96.19\% and 88.94\%, respectively, with only 800 queries. It is demonstrated that current ML methods are often susceptible to systematic black-box attacks. The authors in \cite{8449065} showed that these attacks could drastically degrade classification performance with minimal input perturbations. The use of DL-based algorithms for the wireless physical layer raises serious security and robustness concerns, particularly because these attacks are much more powerful than traditional jamming attempts. The authors in \cite{sadeghi2019physical} explored the creation of physical black-box adversarial attacks and found that the broadcast nature of enemy transmitters significantly increases the block-error rate of communication systems, making these attacks more harmful than traditional jamming. Similarly, Babu et al. in \cite{babu2020physical} demonstrated how adversaries could exploit physical white-box and black-box approaches to compromise DL-based channel decoding with more tremendous success than standard jamming techniques. In conclusion, black-box attacks effectively deceive ML models, decreasing performance and reliability. This challenges the common belief that white-box attacks, with their detailed system knowledge, are inherently more effective. The authors in \cite{majid2023deep} developed a black-box attack method derived from a white-box approach, achieving comparable efficacy with faster execution. Additionally, the authors in \cite{sadeghi2018adversarial} created a white-box attack exploiting specific vulnerabilities in ML algorithms and then created a more efficient black-box version, yielding similar results with reduced data and computational requirements.

\subsection{Adversarial Defense}
In response to adversarial attacks, adversarial defenses were promptly developed. This subsection provides a detailed and insightful overview of adversarial defense strategies.

\begin{figure*}[t]
\includegraphics[width=\linewidth]{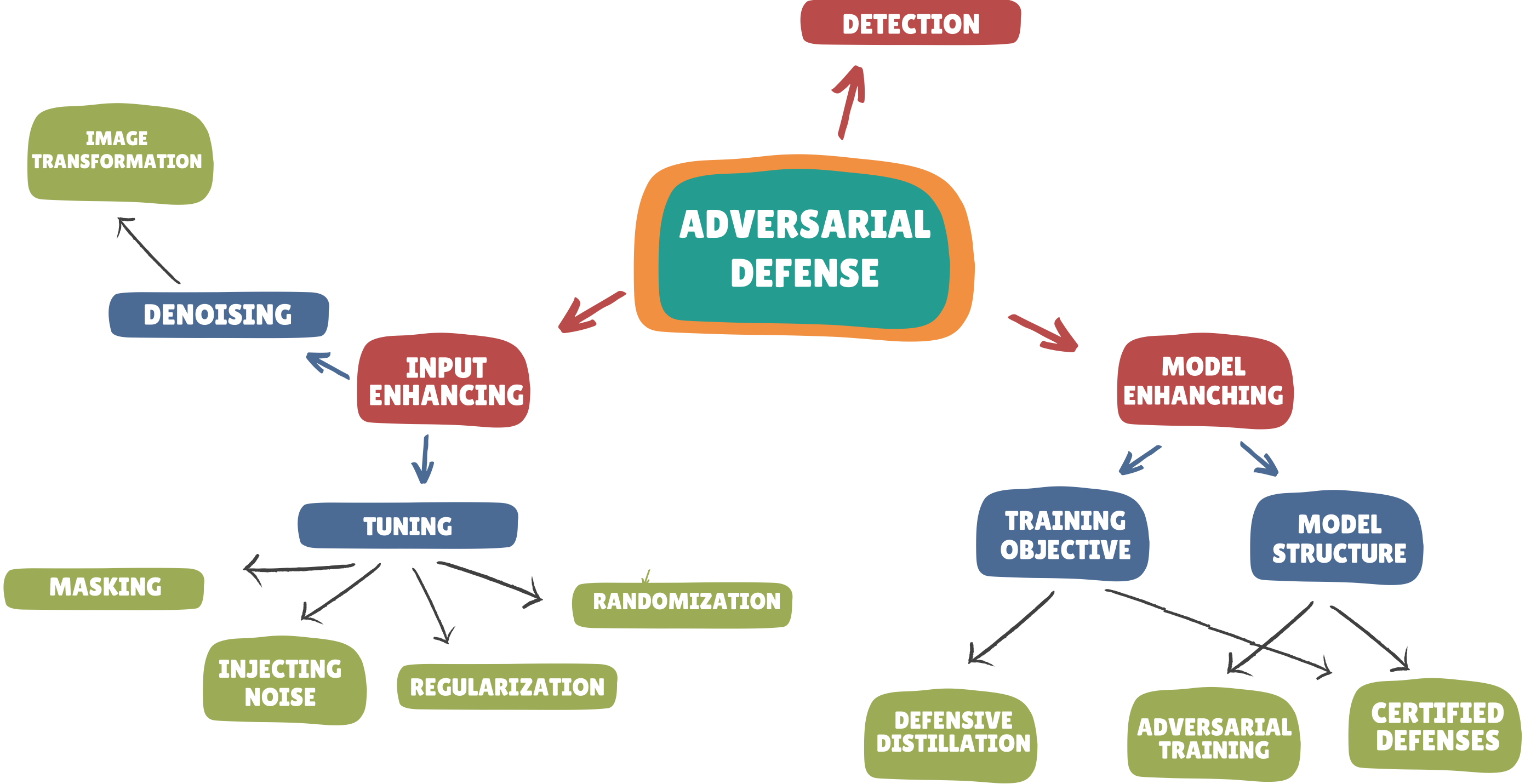}
 \caption{Standard Adversarial Defense Classification}
  \label{sec:def}
\end{figure*}

\subsubsection{Definition}
Adversarial defense refers to techniques and methods designed to protect ML and DL models against adversarial attacks. The primary objective of adversarial defenses is to enhance the robustness and reliability of models, preventing them from being easily deceived by adversaries and ensuring their ability to make precise decisions. Adversarial defenses are particularly crucial when models become targets of adversarial attacks. The key strategies in this domain include adversarial training, input preprocessing, model regularization, and defensive distillation. These strategies aim to strengthen model resilience, enhancing their ability to resist and counter adversarial manipulations.

\subsubsection{Classification}
The fundamental classification of adversarial defense strategies is illustrated in Fig. \ref{sec:def}. The subsequent subsection will elaborate on this classification, providing a more detailed explanation.

Adversarial defense based on model improvement: ML models are highly sensitive to adversarial perturbations. Enhancing their robustness is crucial for countering their sensitivity, leading to increased resilience and reliability.
\begin{itemize}
   \item Adversarial training is employed by both attackers and defenders to enhance model robustness \cite{9964330}. In this method, defense teams expose the model to a blend of perturbed and clean data, acclimatizing it to varied input types. This technique is categorized into two methods: single-step and multi-step adversarial training. Single-step training is effective against one-off, low-complexity attacks such as FGSM and PGD, but it risks catastrophic overfitting, leading to excessive rejection of clean data and significant performance degradation. In contrast, multi-step training is designed to avoid catastrophic overfitting; however, it remains vulnerable to a type of overfitting known as robust overfitting. Despite its utility, adversarial training is a limited defense method. It offers protection against specific attack types, similar to how a vaccine is effective against certain diseases. However, its effectiveness diminishes if the adversary knows the specific adversarial training approach.
   
   \item Defensive distillation is a post-training technique that involves refining a primary model by transferring knowledge from a complex network to a simpler one \cite{9530723}. This technique aims to develop a more refined classifier, better equipped to protect against adversarial attacks that rely on accurate gradient information. However, its effectiveness is challenged by gradient-free or gradient-approximation attack methods. For example, the attack strategy developed in \cite{carlini2017towards} employs norm-restricted additive perturbations capable of completely bypassing defensive distillation. This attack can effectively deceive a defensively distilled network in black-box scenarios, utilizing perturbations derived from an unsecured white-box model. Such attacks significantly undermine the efficacy of defensive distillation in this setting.
  
    \item The use of certified defense methods in adversarial defense has gained popularity \cite{10030859}. This method involves the integration of a noise layer into the baseline model, generating subtle, unexpected perturbations. These perturbations are either smaller than a certain point on the standard input or representative of a specific feature. When injected perturbations remain below the predefined threshold, the model's robustness increases, improving its ability to withstand adversarial attacks.
  
    \item Model randomization, based on integrating noise or random elements into models, challenges adversaries in creating adversarial perturbations \cite{10020163}. By incorporating random factors, the model produces a range of outputs for a single input, thereby obscuring its behavior patterns from attackers. This approach is particularly practical against gradient-based attacks. For example, the model randomization method can make gradient information untrustworthy or fluctuating so that adversaries cannot find an optimal direction to generate the perturbation.
\end{itemize}

 Adversarial defense based on input improvement:
 \begin{itemize} 
     \item Input data serves as a key component in both ML and DL models. The effectiveness of these models depends heavily on the quality of the input data. Tainted or poisoned input can impair the model's ability to make accurate decisions and take appropriate actions. To mitigate the impact of malicious data, various techniques have been employed. One such technique is denoising, where the model is trained to treat adversarial attack data as environmental noise. This approach is often confused with adversarial detection due to the difficulty in distinguishing between the two. However, there is a fundamental difference between adversarial detection and denoising, which is detailed in Table \ref{tab:denoisedetection}.
    
 \item Input reconstruction is a technique that mitigates model vulnerabilities by altering predicted input perturbations through various strategies, ensuring the preservation of data and models. Unlike other methods in this paper that operate during training, its operational phase is during testing. F. Nesti et al. \cite{nesti2021detecting} introduced an innovative input reconstruction method for images, named "defense perturbation," which demonstrated high efficacy as a defense against adversarial attacks. Additionally, the authors of \cite{gupta2019ciidefence} developed CIIDefence, a method that selectively denoises key regions in images to restore normal labels. Alongside pixel-level denoising, they also introduced the HGD method for feature-level denoising. This method involves training a denoising network using a loss function at the feature level, aiming to reduce the disparity between clean and adversarial examples.
 
 \item Input transformation and input reconstruction are fundamentally different in data processing methods, the data's integrity, and the required understanding of the input data type. Input transformation modifies the original data, often resulting in some degree of loss or distortion, in contrast to input reconstruction, which attempts to restore data to its original form. This reconstruction process necessitates an initial understanding of the input data, encompassing its structure, features, and distribution. Conversely, input transformation applies to a broad range of data types without preliminary insights. Y. Qin and C. Yue \cite{9679424} introduced a novel approach to enhance the robustness of input reconstruction defenses by incorporating a key derived from input transformation techniques. Their method, tested in various scenarios including black box, gray box, and white box, proved effective in combating adversarial perturbations and challenging to bypass. While input reconstruction can effectively eliminate noise using a model, well-designed noise can still deceive these models, leading to erroneous outcomes.
 \end{itemize}
 
Adversarial defense based on detection: Adversarial attacks present a significant threat to ML and DL models, exploiting their inherent sensitivities. In response, detection-based adversarial defense techniques have been developed. These techniques focus on identifying and mitigating adversarial perturbations, safeguarding the model from potential damage. Detection-based adversarial defenses are designed to enhance models' robustness against such attacks. Classifying adversarial detection methods presents a challenge owing to the wide variety and complexity of attack types. Nevertheless, these methods can broadly be categorized into two primary types: feature-based adversarial detection and invariant-based adversarial detection.
\begin{itemize}
   \item Feature-based adversarial detection methods employ adversarial and clean data features for detection. For example, \cite{xu2017feature} considers this approach by applying Principal Component Analysis (PCA). This approach combines feature learning with whitening operations to reduce data redundancy. In ML, feature space refers to a domain where each dimension represents a specific data attribute, while feature distance space denotes the metric distance between two data points, measurable via various methods, such as Hamming, Euclidean, Manhattan, and Minkowski distances. In \cite{carrara2018adversarial}, F. Carrara et al. introduced an innovative method for detecting adversarial samples using Euclidean distance within newly defined feature spaces. These spaces represent the relative positioning of a sample to a reference point in the feature space. Their method uniquely encodes the network's activation evolution during forward propagation and distinguishes between the trajectories of genuine and adversarial inputs by embedding internal representations. Preliminary experiments demonstrated the effectiveness of this approach, successfully identifying adversarial inputs targeting the ResNet-50 classifier, pre-trained on the ILSVRC'12 dataset, and generated through various algorithms.
    \item Adversarial detection using the invariant method focuses on identifying attacks in ML models by discerning the discrepancies between adversarial perturbations and clean data. It focuses on learning invariant features capable of classifying data based on semantic content. Recent literature has explored this approach to develop detectors that recognize various attack types, employing invariant activation values and original invariants specific to DNNs. D. Zhou et al. \cite{zhou2021towards} introduced a technique for deriving generalizable invariant features. These features are resilient to attacks and retain the semantic accuracy of classifications. The study proposes a novel method of normalizing attack-invariant features within an encoded space, effectively addressing the bias typically seen between hidden and overt attack types. Their findings indicate enhanced defense efficacy compared to prior methods, particularly against adaptive and novel attack forms. Addressing adversarial attacks in the real world necessitates training models using black or gray box scenarios. In this context, Z. Zheng and P. Hong in \cite{zheng2018robust} introduced an unsupervised learning strategy that models the intrinsic properties of DNN classification independent of prior knowledge of adversary attack techniques. Their study exhibited robust performance against black and gray box attacks, marking a substantial advancement in detection-based adversarial defense.
\end{itemize}
\begin{table}[!t]
\caption{Comparison between 
Adversarial Detection and Adversarial Denoising}
    \centering
    \begin{tabular}{|c|p{3cm}|p{3cm}|}
    \hline
     & Adversarial Detection & Adversarial Denoising\\
    \hline
   Ambition & When considering the data input, the model applied adversarial detection to investigate whether it is an adversarial perturbation. If it is true, the model will reject that data input \cite{mao2023security}.  & When viewing the data input, the model applied adversarial denoising and tried to remove the noise from the input. In that case, it considers the adversarial perturbation like one kind of noise  \cite{salman2019denoising}.\\
    \hline
    Applicability & This technique are often used in the extra defense mechanism and mitigation procedure in the case that the model needs to identify specifically the adversarial perturbation. & Denoising techniques are often used in the preprocessing data phase because it is suitable with the aim of recovering and cleaning the input to the original, besides minimizing the impact of adversarial perturbation.   \\
    \hline
   Outcome & Adversarial detection goal aims to clean the adversarial perturbation and not include other noise. & Adversarial denoising goal aims to clean the noise, including adversarial perturbation.\\
   \hline
    Application  & Autonomous vehicles \cite{amirkhani2022survey}, IoT  \cite{ullah2021framework}, Image processing \cite{liang2018detecting}, etc. & Autonomous vehicles \cite{kloukiniotis2022countering}, IoT  \cite{xiao2022defed}, Image processing \cite{cheng2021pasadena}, etc.\\
    \hline
     \end{tabular}

     \label{tab:denoisedetection}
 \end{table}

\section{ Adversarial Attacks and Defences in 6G network-assisted IoT systems} \label{Sec:AdverDef}

In this section, we examine adversarial attacks and defenses in 6G network-assisted IoT systems. We begin with a background study and review of the existing literature. Next, we propose defensive strategies to protect against adversarial attacks. Finally, we present simulation results to evaluate their effectiveness.
\subsection{Background and research review}

The number of IoT devices has increased dramatically in recent years, leading to significant and rapid growth in research on 6G network-assisted IoT systems. However, as of the writing of this paper, based on our comprehensive understanding, there remains a notable scarcity of literature addressing adversarial attacks and defenses, particularly in the broader scope of future Internet technologies and specifically within the realm of 6G-assisted IoT systems. This observation highlights a critical research gap in the field.

The authors in \cite{9269354} proposed a method that enhances the robustness and sophistication of neural network architectures in the artificial intelligence-enabled IoT (AIoT). This advancement is crucial given that many AI-based massive IoT systems demonstrate vulnerability to adversarial attacks. In this context, adversarial examples for attack and adversarial training for defense were employed. Furthermore, the study proposed a novel multiple-objective gradient optimization approach, which leverages the synergistic effects of adversarial attacks and model delay constraints. This method facilitates a more robust and efficient multi-objective optimization process. Highlighting the relevance in practical applications, automatic vehicles in 6G-assisted IoT networks, which demand low latency and ultra-reliability, are discussed. The immense data processing requirements of numerous sensors in autonomous vehicles necessitate AI support, particularly in neural networks. The article \cite{9390408} introduced an adversarial attack method tailored to unmanned vehicle systems, utilizing incremental learning. By using the FGSM algorithm, the deep-fool algorithm \cite{moosavi2016deepfool}, and the MI-FGSM algorithm \cite{dong2018boosting}, the authors created adversarial examples for attacking the model. The study revealed that incorporating incremental learning into the process significantly enhances the success rate of adversarial attacks. This technique introduces perturbations to clean examples while retaining the model's previously acquired knowledge. The study revealed that this approach results in an 8.43\% increase in attack success rates.

With the growing prominence of reinforcement learning (RL)-enabled massive IoT and the advent of the 6G of wireless networks \cite{ali2021reinforcement}, \cite{li2021drlr}, \cite{xu2021deep}, the necessity for robustness against adversarial attacks has become increasingly critical. Network slicing is one of the key technologies of next-generation radio access networks. It is critical for resource allocation, which has become easier with RL and other learning techniques. In \cite{9984930}, the authors proposed strategies for attacking and defending against network slicing techniques in radio access networks through RL. The authors also employed adversarial ML attacks to generate jamming signals as a defensive measure. Alternatively, federated learning (FL), which involves training algorithms in separate, independent parts without data and model sharing yet still enabling collective model training, shows promise for implementation in 6G networks \cite{yang2022federated}. Similar to RL, FL is also vulnerable to adversarial attacks. To address this issue, the authors of \cite{9288942} developed a robust federated learning architecture in two segments. The first employs dynamic data poisoning via a generative adversarial network (GAN) and a federated generative adversarial network (FedGAN). The second segment uses two counter-GAN algorithms to mitigate aggregation anomalies caused by the GAN network (A3GAN). This attack-defense strategy outperformed existing solutions by 8\%. The vital role of modern wireless networks (5G, 6G, and beyond) in IoT for high data rates and reliability is clear. Techniques combining orthogonal frequency division multiplexing, adaptive modulation (AM), and MIMO aim to achieve high data rates. In their paper, K. Zheng and X. Ma proposed an adversarial attack method for MIMO OFDM with AM, applying a learning-based approach \cite{10038732}. This paper outlined three methods of learning-based adversarial attacks designed to maximize error rates, minimize capacity, or increase outage probability. These methods have proven effective in diminishing the efficiency of the proposed model.

With the growing adoption of 5G/6G networks in various industries, there is an anticipated exponential increase in industrial IoT equipment (IIoTe). Consequently, safeguarding and purifying IIoTe data for finance, healthcare, and digital services applications becomes crucial. Given the significant threat of malware, existing AI-based malware detection methods, though numerous, are still susceptible to adversarial attacks. To combat evasion attacks in the industrial IoT domain, \cite{9817648} introduced a defense strategy against such attacks for malware detection using ML. This approach involves training each classifier within the model on distinct feature subsets. The integrated ML system autonomously counters evasion attacks by employing ensemble-based learning for different classifiers. Their analysis and evaluation showed that this method achieved a 91\% accuracy rate with 14 synthetically generated input features.

\subsection{Simulation}
\subsubsection{Model architecture}
X. YU et al. \cite{9592779} proposed a convolutional auto-encoder for a wireless network-supporting intelligent reflecting surface. Auto-encoder requires a low error rate and a fast processing rate, making DL an ideal fit. However, DL models can become susceptible to adversarial attacks without robust defenses, leading to diminished performance. Even simple attack algorithms, such as those detailed in \ref{alg:FGSM} and \ref{alg:pgd}, can induce auto-encoder errors. We employed the model in \cite{8651357} to assess the impact of adversarial attacks compared to jamming attacks.

\subsubsection{Simulation and result}
\begin{figure}[t]
\includegraphics[width=\linewidth]{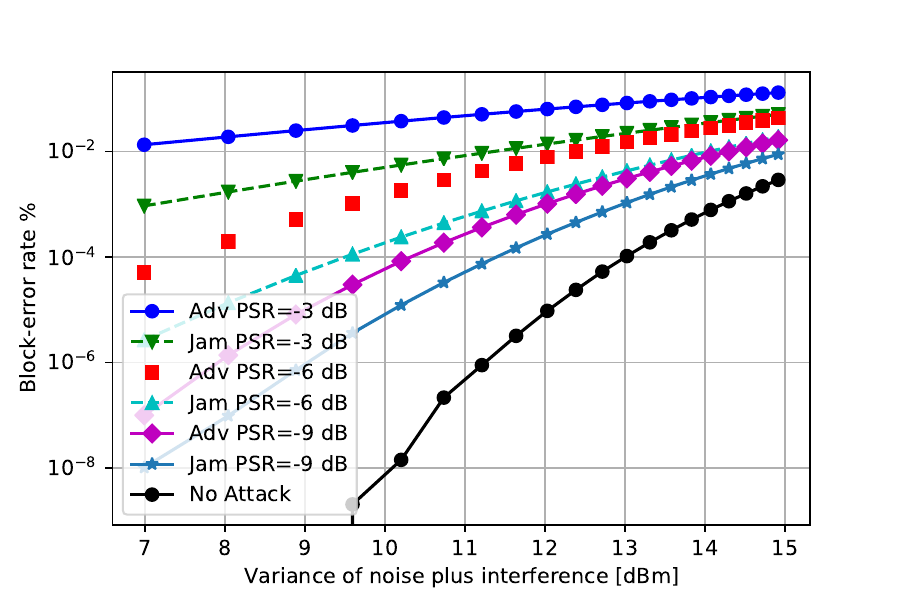}
 \caption{BLER of the auto-encoder system under Jamming and Adversarial attacks.}
  \label{defattac}
  \label{sec:attack}
\end{figure}

In our simulation, the signal power denoted as $E_b$ is set to 50 dBm. We analyze the block error rate (BLER) in scenarios where the noise plus interference variance increases. Specifically, our study investigates the vulnerability of a Multilayer Perceptron (MLP) autoencoder (based on algorithm 1 in \cite{8651357}) to two distinct types of attacks: a Jamming attack using AWGN and an Adversarial attack using the algorithm \ref{alg:FGSM}. To be more precise, the simulation employs a white-box approach for the adversarial attack, a decision driven by the FGSM's requirement for model parameters. Additionally, this approach allows for a straightforward comparison of the effectiveness of adversarial attacks. As illustrated in Fig. \ref{sec:attack}, the results indicate that a basic adversarial attack algorithm, under similar power conditions of received perturbation to the total received signal power, is more effective than a jamming attack employing AWGN.

\section{Research Challenges and Open Issues} 
This section examines research issues regarding the attacks and defense in 6G network-assisted IoT systems. Additionally, we provide Table~\ref{Sec:ResearchChallegnes} that summarizes and reviews adversarial attacks and defenses in 6G network-assisted IoT systems.
\begin{table*}[]
\caption{Open Issues and Challenges in Adversarial Attacks and Defenses in 6G
Network-Assisted IoT Systems} 
\begin{tabular}{llll}
\hline
\multicolumn{2}{|l|}{Challenging} &
  \multicolumn{1}{l|}{Description} &
  \multicolumn{1}{l|}{Key Considerations} \\ \hline
\multicolumn{2}{|l|}{\begin{tabular}[c]{@{}l@{}}Data transmission \\ attack approach\end{tabular}} &
  \multicolumn{1}{l|}{\begin{tabular}[c]{@{}l@{}}Attacking into data transmission leads to the \\ threat that attackers can imitate the ML/DL models \\ models based on collected data.\end{tabular}} &
  \multicolumn{1}{l|}{\begin{tabular}[c]{@{}l@{}}The potential vulnerability of these models\\  to adversarial attacks during training.\end{tabular}} \\ \hline
\multicolumn{2}{|l|}{\begin{tabular}[c]{@{}l@{}}Channel estimation \\ attack approach\end{tabular}} &
  \multicolumn{1}{l|}{\begin{tabular}[c]{@{}l@{}}Channel estimation is essential, but faces \\ performance degradation due to attacks by\\  adversaries.\end{tabular}} &
  \multicolumn{1}{l|}{\begin{tabular}[c]{@{}l@{}}It is crucial to determine whether underwater \\ communication systems are under attack when \\ considering the anomaly BER.\end{tabular}} \\ \hline
\multicolumn{2}{|l|}{\begin{tabular}[c]{@{}l@{}}Centralized attack \\ approach\end{tabular}} &
  \multicolumn{1}{l|}{\begin{tabular}[c]{@{}l@{}}Adversaries attempt to launch attacks on \\ the  CPU of  ML/DL models.\end{tabular}} &
  \multicolumn{1}{l|}{\begin{tabular}[c]{@{}l@{}}Minor disruptions can potentially lead to \\ significant problems due to the vast amount of \\ data processed by the CPU every second.\end{tabular}} \\ \hline
\multicolumn{2}{|l|}{\begin{tabular}[c]{@{}l@{}}Distributed attack \\ approach\end{tabular}} &
  \multicolumn{1}{l|}{\begin{tabular}[c]{@{}l@{}}Adversaries attempt to launch attacks\\  on edge CPU-based ML/DL   models.\end{tabular}} &
  \multicolumn{1}{l|}{\begin{tabular}[c]{@{}l@{}}When refining ECPU based on ML/DL models in \\ a dynamic real-time environment, it is important\\  to minimize potential security vulnerabilities.\end{tabular}} \\ \hline
\multicolumn{2}{|l|}{\begin{tabular}[c]{@{}l@{}}Attack-based \\ power allocation\end{tabular}} &
  \multicolumn{1}{l|}{\begin{tabular}[c]{@{}l@{}}Attackers attempt to deceive ML/DL models \\ that rely on power management units.\end{tabular}} &
  \multicolumn{1}{l|}{\begin{tabular}[c]{@{}l@{}}A combined adversarial attack strategy can be\\  applied to fool ML/DL models, particularly in resource\\  allocation.\end{tabular}} \\ \hline
\multicolumn{2}{|l|}{\begin{tabular}[c]{@{}l@{}}Attack channel \\ coding approach\end{tabular}} &
  \multicolumn{1}{l|}{\begin{tabular}[c]{@{}l@{}}Adversaries aim to decrease performance \\ of joint source-channel coding and others \\ channel coding, which is powered by ML/DL models.\end{tabular}} &
  \multicolumn{1}{l|}{\begin{tabular}[c]{@{}l@{}}The adversaries' ability to create black box\\ attacks for high-performance blind attacks by \\ creating diverse white box models based on\\  input and output parameters.\end{tabular}} \\ \hline
\multicolumn{1}{|l|}{\multirow{5}{*}{\begin{tabular}[c]{@{}l@{}}Attack into \\ cutting-edge\\ technologies \\ in 6G\end{tabular}}} &
  \multicolumn{1}{l|}{\begin{tabular}[c]{@{}l@{}}Reconfigurable\\  Intelligent \\ Surfaces\end{tabular}} &
  \multicolumn{1}{l|}{\begin{tabular}[c]{@{}l@{}}ML/DL models assists RIS in having high performance\\  in 6G, but also makes it vulnerable to attacks.\end{tabular}} &
  \multicolumn{1}{l|}{\begin{tabular}[c]{@{}l@{}}It is crucial to ensure the security and robustness of\\  RIS models when applying modern learning\\ techniques under the threat of adversarial attacks.\end{tabular}} \\ \cline{2-4} 
\multicolumn{1}{|l|}{} &
  \multicolumn{1}{l|}{\begin{tabular}[c]{@{}l@{}}Satellite-terrestrial\\  communications\end{tabular}} &
  \multicolumn{1}{l|}{\begin{tabular}[c]{@{}l@{}}Attackers aim to degrade the performance \\ of satellite-terrestrial communication design  \\ based on ML/DL models.\end{tabular}} &
  \multicolumn{1}{l|}{\begin{tabular}[c]{@{}l@{}}Although powered by ML/DL models, these techniques\\ also leave satellite-terrestrial communication \\ susceptible to new physical layer security \\ vulnerabilities.\end{tabular}} \\ \cline{2-4} 
\multicolumn{1}{|l|}{} &
  \multicolumn{1}{l|}{\begin{tabular}[c]{@{}l@{}}Massive MIMO/\\ Cell-Free \\ Massive MIMO\end{tabular}} &
  \multicolumn{1}{l|}{\begin{tabular}[c]{@{}l@{}}New networking features have been observed in \\ Massive MIMO and Cell-Free Massive MIMO, \\ which opens doors for attackers to exploit\\  vulnerabilities in the physical layer of security.\end{tabular}} &
  \multicolumn{1}{l|}{\begin{tabular}[c]{@{}l@{}}Massive MIMO and Cell-Free Massive MIMO have\\ created new features in physical layer security due to\\massive antennas array. Imperfect channel state \\information  affects uplink and downlink data \\transmission, leading to the need for new attack strategies.\end{tabular}} \\ \cline{2-4} 
\multicolumn{1}{|l|}{} &
  \multicolumn{1}{l|}{Metaverse} &
  \multicolumn{1}{l|}{\begin{tabular}[c]{@{}l@{}}Attackers attempt to breach edge/cloud \\ computing servers based on ML/DL models and \\ focusing on the reality of transformed data.\end{tabular}} &
  \multicolumn{1}{l|}{\begin{tabular}[c]{@{}l@{}}Although DRL algorithms effectively optimize\\ offloading performance for Metaverse's \\ offloading process and  Metaverse's transferring \\ process, adversarial attacks can decrease models' \\performance by using a data poisoning approach.\end{tabular}} \\ \cline{2-4} 
\multicolumn{1}{|l|}{} &
  \multicolumn{1}{l|}{\begin{tabular}[c]{@{}l@{}}Semantic \\ communications\end{tabular}} &
  \multicolumn{1}{l|}{\begin{tabular}[c]{@{}l@{}}During data transmission, adversaries can \\ easily target and exclude redundant data.\end{tabular}} &
  \multicolumn{1}{l|}{\begin{tabular}[c]{@{}l@{}}It is crucial to enhance securely transmit data as \\ adversarial attacks are more dangerous in this phase.\end{tabular}} \\ \hline
 &
   &
   &
   \\
 &
   &
   &
  
\end{tabular}
\end{table*}
\label{Sec:ResearchChallegnes}
\subsection{Attack into transmission phase}
Data transmission occurs through sending and receiving over a communication channel, which can be either wired or wireless. In the context of future IoT systems, the transmission protocol is supported by promising technologies, including semantic, quantum, and satellite communications. Data serves as the fundamental element of communication, and the effectiveness of data transmission plays a crucial role. Attacks utilizing datasets primarily focus on manipulating the input of a learning model. Data are transmitted in the context of 6G-assisted IoT networks, it is vulnerable to interference by malicious attackers through stealing information or compromising trustworthiness. Note that channel estimation holds significant importance in a communication system. This work aims to enhance signal quality by accurately estimating channel characteristics. This, in turn, improves receiver performance in terms of signal detection, demodulation, and decoding. Moreover, channel estimation aids the model in mitigating interference by estimating channel parameters and facilitating the differentiation between desired and interfering signals.
\subsubsection{Data transmission attack approach} 
Adversaries often manipulate a model's decision-making process through adversarial attacks, such as inference and trojan attacks. Notably, data poisoning \cite{wang2021robust} and evasion attacks \cite{sagduyu2019iot} are prominent methods of corrupting data. Particularly in wireless data transmission, adversaries can observe and replicate DL models, practicing in both white-box and black-box scenarios. They aim to approximate the target model's parameters and identify the most effective attack algorithm. In underwater communications, where data transmission experiences higher path loss than air-borne transmission, the block error rate (BLER) may increase, rendering models more susceptible to adversarial attacks. For example, W. Zhang et al. \cite{9694507} demonstrated using CNNs, DNNs, and hybrid neural networks (RDNNs) for modulating underwater acoustic signals. The RDNN proposed by the authors achieved higher accuracy and lower time processing. Nevertheless, the study did not address the potential vulnerability of these models to adversarial attacks during training. Generally, if adversarial attacks compromise a model, the outcomes can be detrimental.

\subsubsection{Channel estimation attack approach} 
The critical role of channel estimation in communication systems, coupled with the advantages of ML and DL in this domain, makes these systems prime targets for adversarial attacks. These attacks, aimed at degrading channel estimation performance, occur during the data collection and training phases of the model. In \cite{bahramali2021robust}, the authors explored the use of white-box and black-box attacks on OFDM channel estimation, employing algorithms \ref{al:channelestimate} and \ref{al:gan_noise_regularizer}. Furthermore, in underwater wireless optical communication, which is essential for high data rate services, various environmental factors, such as unknown objects, temperature fluctuations, and air bubbles in the water challenge channel estimation \cite{9302692}. However, a pressing concern arises when considering how a DL model might confuse adversarial perturbations with normal interference signals. Such confusion could lead to reduced accuracy and erroneous predictions during online training. The authors need to address the impact of both regular interference and adversarial perturbations in underwater environments.

\subsubsection{Defense approach} 
In defending against data transmission attacks, the authors in \cite{steinhardt2017certified} employed certified defenses to counteract data poisoning. Additionally, combined methods like denoising can provide enhanced protection. For channel estimation, it is evident that selecting a single defense strategy that performs optimally against various types of attack is a complex challenge. However, in some cases, as demonstrated by the authors in \cite{catak2022defensive}, the distillation defense method, the same approach we discussed, was employed to reduce adversarial perturbation effectively. Data transmission and channel estimation are integral components of wireless communication systems. As new adversarial attack methods emerge, it is crucial for defensive strategies to evolve rapidly. Successful attacks on these systems or models, whether in real or virtual environments, can cause significant damage, leading to loss of customer trust and negative experiences.
\begin{algorithm}
    \caption{Generating universal adversarial attack using perturbation generator model}
    \label{al:channelestimate}
    \begin{algorithmic}[1]
        \State \textbf{Init:}
            \State $\mathcal{D}$ represents the data used for training an adversarial model.
            \State $f$ denotes a model based on deep neural networks (DNNs).
            \State $y$ represents the input used for training.
            \State $\mathcal{L}_f$ is the loss function based on DNNs.
            \State $\mathcal{M}$ refers to the function that remaps the domain.
            \State $\mathcal{R}$ represents the function used for domain regularizations.
            \State $G(z)$ denotes the initialization of the model for blind adversarial perturbations.
            \State Parameters $(\theta_G)$ represents the parameters associated with the blind adversarial perturbation model.
            \State $T$ represents the number of epochs (training iterations).
        \For{$\text{epoch } t \in \{1, \ldots, T\}$}
            \For{all mini-batch $b_i$ in $\mathcal{D}$}
                \State $z \sim \text{Uniform}$
                \State Rotate $\mathcal{M}(y, G(z))$ based on the channel phase shift
                \State $J = -\left(\frac{1}{|b_i|} \sum_{x \in b_i} \ell(f(\mathcal{M}(y, G(z))), f(x))\right) + R(G(z))$
                \State Update $G$ to minimize $J$
            \EndFor
        \EndFor
        \State \textbf{return} $G$
    \end{algorithmic}
\end{algorithm}
\begin{algorithm}
    \caption{GAN-based Noise Regularizer}
    \label{al:gan_noise_regularizer}
    \begin{algorithmic}[1]
        \State $\mathcal{D}$ represents the training data.
        \State $f$ denotes a model based on DNNs.
        \State $G$ represents the perturbation generator model (PGM).
        \State $D$ represents the discriminator model.
        \State $\mu$, $\sigma^2$ represent the parameters for the desired Gaussian distribution.
        \For{$t \in \{1, 2, \ldots, T\}$}
            \State $z_0 \sim \text{Gaussian}(\mu, \sigma^2)$
            \State $z \sim \text{Uniform}$
            \State Train $D$ on $G(z)$ with label 1 and $z_0$ with label 0
            \State Train $G$ on $D^S$ using regularizer $\mathcal{R}$
        \EndFor
        \State \textbf{return} $G$
    \end{algorithmic}
\end{algorithm}

\subsection{Centralized and distributed attacks} 
In initiating attacks, adversaries face a strategic choice: invest in costly equipment to target a centralized system or acquire multiple low-cost devices for a coordinated, distributed attack. The effectiveness of these approaches varies depending on the scenario. Specifically, within the context of FL, pivotal for enhancing 6G-supported IoT applications, authors in \cite{zhang2022federated} investigate the challenges, methods, and future directions of centralized and distributed attack approaches. A prominent challenge identified is limited resources and bandwidth constraints, where FL can significantly improve. However, user privacy in FL systems necessitates robust defense mechanisms against adversarial attacks. Adversaries could exploit FL to collect data despite technological advancements. In the following section, we consider two types of adversarial attacks to explore the vulnerability of FL-assisted IoT systems.

 \subsubsection{Centralized attack approach}
A centralized processing unit (CPU) has a better computational capacity higher performance, and is easier to manage and maintain than a distributed central processing unit (DCPU). However, its high cost, substantial size, and significant electricity consumption render it less efficient, posing economic trade-offs when selecting between a CPU and an ECPU. CPUs are commonly employed in industries that provide big data processing and cloud services. Therefore, the vulnerability of these companies' CPU-integrated DL systems is a critical concern. Minor disruptions can potentially escalate into significant problems due to the vast amount of data the CPU processes every second. Controlling adversarial perturbations becomes particularly challenging if they infiltrate training data. In such scenarios, data poisoning emerges as a potential and concerning threat.

 \subsubsection{Distributed attack approach} Attack approach: 
 In the context of massive IoT, data streams are directed back to the ECPU, which offers reduced latency and a lower block error rate despite its limited data processing capacity. Leveraging DL significantly enhances this setup, decreasing computation time, conserving resources, and simplifying management. For example, deploying a local 6G network in automated manufacturing can facilitate URLLC for massive IoT systems, enabling efficient data management. Short-packet communications augmented with energy harvesting become increasingly effective when integrated with intelligent data processing units. In a dynamic real-time environment, updating and refining the DL models for the ECPU frequently is essential. However, it introduces potential security vulnerabilities. For instance, adversaries might exploit these opportunities to deploy various adversarial attack methods, aiming to deceive ML models and compromise the system.
 

\subsubsection{Defense approach} 
 In \cite{segura2021centralized}, the authors proposed a detector suitable for defense against centralized and distributed attacks. Instead of ML or DL, the detector operates effectively without extensive training. As a result, the detector achieves high performance without training delay. A significant advantage of this approach is its inherent resilience to initial attack attempts; it reduces the likelihood of the model being compromised during the initial stages of defense training. To further bolster the robustness of this proposed model against centralized and distributed attacks, we recommend incorporating ML and DL techniques, such as adversarial training. This could lead to a hybrid defensive strategy, enhancing overall system resilience. While both centralized and distributed attacks can be effective, attackers need to consider the financial and temporal costs involved. Similarly, defenders must also consider these factors to select an efficient defense strategy that effectively addresses financial and temporal constraints.

\subsection{Attack into resource allocation}
While DL is effective in power control for wireless communication applications, it is also vulnerable to adversarial attacks. This subsection introduces attack-based and defense-based resource allocation methods in 6G-assisted IoT systems.

\subsubsection{Attack-based power allocation}
Power allocation, a crucial process in distributing the power budget among users based on channel state information, plays a pivotal role in IoT systems where energy saving is vital for environmental protection, reduced electricity consumption, and economic efficiency. DL significantly optimizes resource allocation within massive MIMO technology, the backbone of IoT systems. However, attackers can easily target this process. We will explore various strategies for attacking and defending power allocation mechanisms. Due to the importance of power allocation, several research articles have been published. For example, the authors in \cite{santos2021universal} proposed a universal adversarial attack targeting DL-based neural networks in massive MIMO systems for power allocation. They implemented a universal adversarial perturbation (UAP) using a minimum perturbation algorithm (Algorithm \ref{minimumpertubation}) to minimize interference and maximize attack efficacy. Additionally, they applied accumulative perturbation and PCA-based methods with $||L||_\infty$ norm for crafting UAPs. By leveraging an optimized white-box attack strategy (Algorithm \ref{optimizedwhiteboxattack}), their approach demonstrated superior effectiveness compared to traditional white-box attacks using FGSM (Algorithm \ref{alg:FGSM}) and PGD (Algorithm \ref{alg:pgd}). A combined adversarial attack can be applied to fool DL models, particularly in resource allocation. Such attacks can significantly impact the energy efficiency of 6G-enabled AIoT systems, leading to increased production costs and loss of consumer trust.

\begin{algorithm}[t]
\caption{Minimum Perturbation Algorithm}
\label{minimumpertubation}
\begin{algorithmic}[1]
\Procedure{MINPERTURBATION}{$f, x, \epsilon, P_{\max}$}
    \State \textbf{Input:} $x, P_{\max}, f(\cdot, \theta), \Delta_{\max}, \epsilon_{\text{acc}}, I_{\max}$
    \State \textbf{Output:} $\delta$
    \State Initialize: $\epsilon_{\text{max}} = \Delta_{\max}$, $\epsilon_{\text{min}} = 0$, $I = 0$
    \While{$(\epsilon_{\text{max}} - \epsilon_{\text{min}} > \epsilon_{\text{acc}})$ and $(I < I_{\text{max}})$}
        \State $I = I + 1$
        \State $\epsilon = (\epsilon_{\text{max}} + \epsilon_{\text{min}})/2$
        \State $x_{\text{adv}} = x + \epsilon \cdot \text{sign}(\nabla_x L_j (f(x)))$
        \If{$\sum_{k=1}^{K} \hat{\rho}_{jk} < P_{\max}$}
            \State $\epsilon_{\text{min}} = \epsilon$
        \Else
            \State $\epsilon_{\text{max}} = \epsilon$
        \EndIf
    \EndWhile
    \State $\delta = \epsilon_{\text{max}} \cdot \text{sign}(\nabla_x L_j (f(x)))$
    \State \textbf{return} $\delta$
\EndProcedure
\end{algorithmic}
\end{algorithm}
\begin{algorithm}[t]
\caption{Optimized White-box Attack}
\label{optimizedwhiteboxattack}
\begin{algorithmic}[1]
\Procedure{WhiteBoxAttack}{$x, P_{\max}, f(\cdot, \theta), \epsilon, I_{\max}$}
    \State \textbf{Input:} $x, P_{\max}, f(\cdot, \theta), \epsilon, I_{\max}$
    \State \textbf{Output:} $\eta$
    \State Initialize: $\eta = 0.2K\ell$
    \For{$i$ \textbf{in} range($I_{\max}$)}
        \State $x_{\text{adv}} = x + \eta$
        \State $x_{\text{adv}} = \text{clip}(x_{\text{adv}}, x - \epsilon, x + \epsilon)$
        \State Obtain $\arg \min_{\eta} -L_j(f(x_{\text{adv}}))$
        \If{$\sum_{k=1}^{K} \hat{\rho}_{jk} > P_{\max}$}
            \State \textbf{break}
        \EndIf
    \EndFor
    \State \textbf{return} $\eta$
\EndProcedure
\end{algorithmic}
\end{algorithm}

\subsubsection{Attack channel coding approach} 
DL has significantly advanced wireless communications, especially in joint source-channel coding (JSCC). This progress is not confined to JSCC at specific signal-to-noise ratios (SNRs) but extends to a diverse range of SNRs. The authors in \cite{9438648} proposed a novel DL-based approach to source-channel coding in wireless image transmission. The authors demonstrated that their study is the first to examine JSCC across a range of SNRs, marking a pioneering approach in this field. The results indicate that their approach offers greater adaptability, robustness, and versatility than traditional approaches. However, a notable limitation of DL, as highlighted in the study, is its susceptibility to adversarial attacks. Therefore, DL-based JSCC may underperform compared to traditional JSCC, necessitating careful consideration of adversarial attacks in such implementations. Concurrently, channel coding parameters play a significant role in the effectiveness of the coding scheme. In this context, S. Dehdashtian et al. presented an advanced DL approach for blind detection of channel coding parameters. This approach demonstrated superior performance across various coding schemes, including turbo, polar, LDPC, and convolutional codes, surpassing traditional approaches. It exhibits strong robustness against multi-path fading and operates independently of prior SNR and channel state knowledge, making it highly effective for accurately detecting coding parameters. However, this model type is vulnerable to data poisoning attacks during its training phase. Furthermore, adversaries can exploit the model's input and output to create diverse white-box scenarios, effectively treating the model as a black box to identify the most effective attack strategy.

\subsubsection{Defense approach}

Various methods have been proposed to improve the robustness of AIoT-assisted 6G systems against adversarial attacks. R. Sahay et al. \cite{sahay2023defending} proposed a defense mechanism utilizing adversarial denoising. The results show that this method effectively mitigates the impact of adversarial attacks on power allocation in massive MIMO systems, particularly against attacks developed using FGSM and PGD. However, its performance could be less effective against more sophisticated attack methods, such as C$\&$W. This limitation highlights the necessity for developing more sophisticated and, possibly, integrated defense strategies. While numerous defenses are available to enhance the robustness of DL models employing auto-encoders, reliance on outdated methods is precarious due to the evolving nature of adversarial attacks. Adversaries operating in white-box scenarios can identify and employ the most effective attack strategies against these models. Auto-encoder protection remains a critical area of research, requiring further exploration in future studies.

\subsection{Attack into cutting-edge technologies  in 6G}
\subsubsection{Reconfigurable Intelligent Surfaces} 
In the rapid growth of  IoT networks, reconfigurable intelligent surface (RIS) and its cutting-edge subfields, including simultaneously transmitting and reflecting RIS (STAR-RIS), offer transformative possibilities for wireless communication \cite{waqs1,waqs2}. However, the presence of RISs also introduces emerging security challenges. RIS, with the backing of DL and ML, can collaborate with many advanced technologies to improve the performance of the sixth-generation networks. For instance, the authors in \cite{10136735} proposed an auto-encoder replacing the RIS-assisted MIMO systems with the controlled phase shifts. In particular, the result showed that the auto-encoder decreased the symbol error rate and boosted system performance. However, the  RIS-assisted system did not consider the possibility of being damaged in the training and testing phases. If one studies adversarial attacks, the auto-encoder will enhance the robustness when applying defense techniques to face the dangers. To optimize the phase shift coefficients, modern learning-based neural networks are a good option to perform based on real data sets. B. Sagir et al. \cite{10025789} presented IoT networks with the support of the RISs for cooperative communications using DL. The results demonstrated promising applications of learning-based methods for RIS-assisted models with low-cost algorithm designs. Nonetheless, this work did not consider the risk of adversarial attacks. It is crucial to ensure the security and robustness of RIS models when applying modern learning techniques under the threat of adversarial attacks.

\subsubsection{Massive MIMO/Cell-Free Massive MIMO} A large number of antennas equipped at a compact array or in a distributed manner has manifested their roles in improving spectral and energy efficiency for the entire networks \cite{van2021reconfigurable}. These technologies allow for more efficient use of the available spectrum and enhance wireless networks' reliability by exploiting linear signal processing techniques such as maximum ratio or zero forcing to achieve a near-optimal solution \cite{elhoushy2021cell}. Due to many antennas with a rich resource of randomness, Massive MIMO, and Cell-Free Massive MIMO can effectively cope with fast-fading channels that ergodic performance metrics become relevant to evaluate system performance. Thanks to increasing freedom, the networks can simultaneously serve many users by sharing the same time and frequency resources. Moreover, an overhead dedicated to channel estimation is not neglectable in many scenarios, especially in high-mobility networks with short coherence time. New networking features observed in Massive MIMO and Cell-Free Massive MIMO open room for challenging problems in the physical layer security. Practical considerations such as imperfect channel state information influencing uplink and downlink data transmission phases call for new attack strategies. This is because pilot reuse makes different patterns among the users, and coherence interference reduces network performance dramatically. Furthermore, several questions, including whether or not we should attack both the pilot training and data transmission phases, create challenging research issues to maximize the network degradation as the attackers desire. Besides, long-term network management in Massive MIMO and Cell-Free Massive MIMO with many objective functions requires attackers to design attacking algorithms that are functions of channel statistics.
\subsubsection{Satellite-terrestrial communications} According to the 3GPP RP-230706  \cite{10173867}, satellites should be integrated into 5G advanced and 6G systems. Note that satellite communications can provide telecommunications services worldwide. In contrast, terrestrial networks, including base stations and access points, cover a limited region effectively as urban. The integrated networks open many challenging issues, which call for the applications of optimization theory, artificial intelligence, algorithm designs, and network technologies to enhance system performance and communication reliability. There are several different categories of satellites, including low earth orbit (LEO), medium earth orbit (MEO), and geostationary orbit (GEO). Among them, the most promising infrastructure for the global IoT network is LEO satellites moving around our planet about 160-1500 kilometers from the ground. LEO satellites are often exploited to support military, remote sensing, and scientific research applications. On the other hand, the field of the physical layer in satellite-terrestrial communications faces challenging issues, which can be handled by the support of DL. The work in \cite{9826890} has applied a DNN to collect downlink channel state information and to perform the beamforming design with the support of an LEO satellite. Furthermore, the authors in \cite{9171480} showed that multiple-beam satellites are vulnerable to jamming attacks. It means that one can use adversarial attacks to damage DL-supported satellite communications.
\subsubsection{Metaverse} The definition of ``Metaverse" was introduced in 1992 in a science fiction novel by Neal Stephenson. Regarding network communications, Metaverse implies a virtual environment that integrates the physical and digital worlds. Advanced radio architectures stimulate this technology to boost spectral and energy efficiency. It offers users immersive experiences and new ways of interaction but also poses new cybersecurity challenges that require a thorough analysis. As aggressive requests for high throughput from many users, a Metaverse system must process a huge amount of data. Hence, offloading parts of data into edge/cloud computing is paramount to reduce the computational complexity of the core network. However, data received from the physical world is often not constant, meaning that the Metaverse system needs dynamic offloading methods. Specifically, deep reinforcement learning (DRL) algorithms were used \cite{10122221} to offload dynamic data. Even though DRL algorithms were demonstrated to be suitable for optimizing offloading performance, the adversarial attack can make that offloading model increase the performance by using a data poisoning approach. Besides computing techniques, digital twins are an early version of constructing the ultimate Metaverse system. Digital twins required multiple properties \cite{vaezi2022digital}. If edge/cloud computing presents the promptness property, image classification can be considered the similarity property. When transferring an image into the Metaverse system, its content must be moderated by utilizing, for example, DL. Due to community standards such as age inappropriateness, sensitive data must be detected and ignored. As mentioned, various ways that force DL models to make wrong decisions exist. Consequently, these DL models will decrease the prediction performance by exploiting adversarial attacks.
\subsubsection{Semantic communications (SemCom)} 
This novel paradigm goes beyond the traditional Shannon’s information theory and enables the exchange of meaningful information between humans and machines in future networks \cite{9955312}. SemCom is effective and requires less processing time because it transmits only selected portions of information rather than all the data. Nevertheless, SemCom is one of the most vulnerable systems facing adversarial attacks. Adversaries would be easier to target in the data transmission phase. Generally, they do not need to focus on redundant data, which increases the attacking time and complexity. For example, in image processing, particularly the license plate recognition system, the camera takes a picture that includes significant data such as license plate and other redundant data. The artificial intelligent model will identify the number of license plates and transmit them to the cloud-based SenCom center. That time is the opportunity for an attack to appear using the centralized attack on the significant data. The model may have been fooled and made the wrong decision.

\section{Conclusion}
\label{sec:conclusion}
This paper investigates adversarial attacks and defenses for 6G network-assisted IoT systems, covering various aspects. We analyze and categorize advanced attack and defense strategies, highlighting their impact on IoT security. We also explore the challenges and opportunities of security in the 6G era, focusing on key areas such as data transmission, device density, resource allocation, and the integration of advanced technologies driven by ML and DL. Furthermore, we emphasize the critical need for robust and adaptive adversarial defense strategies, employing modern methods such as anomaly detection and adversary training, to protect 6G networks from emerging threats. In conclusion, this paper provides a comprehensive guide for future research and development on securing 6G network-assisted IoT systems against adversarial attacks. As the 6G landscape evolves, the insights offered in this paper will be valuable, ensuring that the next generation of wireless communication technology remains secure and resilient.


\bibliographystyle{IEEEtran}
\bibliography{IEEE}
\end{document}